\documentclass[%
 reprint,
 amsmath,amssymb,
 aps,
]{revtex4-2}

\usepackage{graphicx}% Include figure files
\usepackage{dcolumn}% Align table columns on decimal point
\usepackage{bm}% bold math
\usepackage{xcolor}
\UseRawInputEncoding
\begin{document}

\preprint{APS/123-QED}

\title{Structural transformation of dusty plasma crystal in DC discharge plasma by changing confinement ring bias}

\author{S. Jaiswal}
%\altaffiliation[Also at:~]{Department of Physics and Astronomy, Eastern Michigan University, Ypsilanti MI 48197, USA}
\affiliation{Department of Physics, Indian Institute of Science Education and Research, Dr. Homi Bhabha Road, Pune 411008, India}
\email{surabhijaiswal73@gmail.com}

\author{Connor Belt}
\affiliation{Department of Physics and Astronomy, Eastern Michigan University, Ypsilanti MI 48197, USA}

\author{Anton Kananovich}
\affiliation{Department of Physics and Astronomy, Appalachian State University, Boone, North Carolina 28608, USA}

\author{E. M. Aguirre}
\affiliation{Rogue Space Systems Corporation, Laconia, NH 03246 USA}

\date{\today}

\begin{abstract}
We report the experimental study of the structural transition of a stable complex plasma crystal to a solid-liquid phase coexistence by the controlled adjustment of the confinement potential, while keeping all other parameters constant. The experiments are carried out in a tabletop Linear Dusty Plasma Experimental (LDPEx) device which consists of a circular powered electrode and an extended grounded cathode plate. A stationary crystal of melamine formaldehyde particles is formed in a background of Argon plasma inside a confining ring that is isolated to the cathode by a ceramic cover. The stable crystal structure breaks in the core region and transitions to a coexistent state by carefully changing the confining potential, thereby modifying the sheath structure. The transition is confirmed by evaluating the variation in different characteristic parameters such as the pair correlation function, local bond order parameter, and dust kinetic temperature as a function of confining bias potential. It is found that melting in the core is due to the onset of dust fluctuations in the layers beneath the topmost layer, which grow in amplitude as the confining bias potential is reduced below a threshold value. The present technique of changing confinement provides a unique feature to study structural transitions of plasma crystals without affecting the overall plasma parameters.
\end{abstract}

%\keywords{Suggested keywords}%Use showkeys class option if keyword
                              %display desired
\maketitle

\section{\label{sec:level1}Introduction}
``Dusty'' or ``complex'' plasmas consist of the usual combination of electrons, ions, and neutral atoms with the addition of charged, dust particulates of size ranging from tens of nanometers to tens of micrometers. The dynamics of each component exhibits different time and length scales. In the plasma environment, these heavy dust particles can acquire a negative charge of the order of Z$_d$ $\approx 10^4$-$10^5$ elementary charges due to the collection of more electrons than ions, which adds richness to the collective dynamics of the system. In a laboratory environment, these particles levitate near the sheath boundary by balancing the electrostatic force due to the sheath electric field and gravity and interact via a screened Yukawa potential. The consequential rich dynamics exhibits interesting phase and structural changes in the system. The phase of the dust system can be determined by a coupling parameter $\Gamma$, which is the ratio of the interparticle Coulomb potential energy to the dust thermal energy. When $\Gamma$ exceeds a critical value, the system forms a so-called Coulomb crystal; a regular lattice structure. The control over the coupling strength of the system makes dusty plasma an excellent platform for studying the phenomenon of phase transitions, transport properties of the strongly coupled system, and other related topics. This has great applicability in areas such as statistical mechanics \cite{statistical}, soft condensed matter \cite{condensed_matter}, and colloidal suspensions \cite{hamaguchi}. The spatial and time scales of the particle motion allow the particles to be visualized using laser illumination and high-speed cameras. Further, the weak frictional damping enables the measurement of the dynamics and kinetics of individual particles. Such a diagnostic convenience of dusty plasma motivated a great deal of research related to crystal formation, phase transitions of two and three dimensional crystals \cite{melzer_crystal_1996, jaiswal_melting, schweigert2,numkin}, re-crystallization \cite{christina}, instabilities \cite{lenaic, meyer}, magnetic field effect on the phase transition \cite{jaiswal_magnetic}, heat transport \cite{nonumura_heat, fortov_heat} and much more.

\par Phase coexistence is a phenomena deeply important to multiple disciplines which has attracted attention in recent years \cite{phase_tutorial}. Defining phase coexistence in non-equilibrium systems is difficult, but it is generally viewed as the existence of multiple macroscopic phases separated by an interface of microscopic thickness \cite{dickman}. Phase coexistence has recently been studied in grains \cite{grains}, colloidal fluids \cite{colloid_coexistence}, and active Brownian particles \cite{brownian}. There has been much interest in active matter showing interesting non-equilibrium behavior for a variety of technological purposes \cite{active_matter}. The Gibbs phase rule relates the degrees of freedom and the number of phases in ``pVT'' (pressure, volume, temperature) states. Simultaneous phase coexistence leads to the well known triple point in components such as water. However, recent studies of phase coexistence suggest that violations/modifications of this rule are possible leading to a quadruple point \cite{liquid_gas}, or in colloidal systems a quintuple point \cite{quintuple}. 

\par Recent work in dusty plasma have also realized a form of phase coexistence \cite{hariprasad_phase,swarnima_pop} but this has received limited exposure so far. On the contrary, there have been several studies devoted to understanding the phase transition of melting in dusty plasma crystals caused by various mechanisms. These studies have identified the mode-coupling instability (MCI) \cite{lenaic,coudel_review}, the Schweigert instability \cite{schweigert1,schweigert2,schweigert3}, charge fluctuations \cite{jaiswal_melting}, magnetic field induced shear \cite{jaiswal_magnetic}, and laser melting \cite{laser_melting}. Some of these experiments showed signatures of phase coexistence as a transient state without a clearly discernible sharp interface between the macroscopic states. For instance, once the MCI is triggered the melting proceeds from the center of the crystal to the entire monolayer \cite{mci_propagation,rocker_melting}, showing transient phase coexistence. Recent molecular dynamic simulations showed that non-uniformity of monolayer properties can create a state of phase coexistence in a dusty plasma system \cite{nikolaev_coexistence}. For a monolayer with plasma wakes, a central fluid core and solid periphery was demonstrated. Hariprasad \textit{et al.} presented ``self-sustained'' phase coexistence in a DC system for a 3D dust crystal with a liquid central region and a peripheral 2D crystalline state by varying the neutral pressure \cite{hariprasad_phase}. The observed coexistence was driven by the non-reciprocal wake effect, which ultimately gives rise to an instability that is driven locally in the system. Later, Swarnima \textit{et al.}\cite{swarnima_pop} conducted experiments in a DC monolayer plasma on the transition of a 2D crystal to a self-sustained 2D two-phase coexistence state. They identified the nature of the instability from the horizontal oscillation of particles as being the Schweigert instability that occurs prior to the coexistence. In both 3D and 2D experiments they varied discharge parameters which affect the background plasma condition. Theories of melting and subsequent phase coexistence are developed for bilayer crystals (Schweigert instability) or monolayer crystals (MCI). However, the analysis required to definitively confirm the mechanism is quite complex. With regard to 3D systems, the question of whether the phenomena are driven solely by an instability or are due to the formation of multilayered systems under the effect of an electrostatic trap are still not clear and need more precise experiments. 

\par In this paper, we have explored the phenomenon of phase coexistence by carefully adjusting the potential of a confining ring without altering the discharge and global plasma properties. We have observed the transition of a stable crystalline structure into a two-phase coexistence state when the confining bias potential is reduced below a threshold value. The two distinct macroscopic states with different temperatures exist simultaneously and are separated by a sharp circular interface which stays constant in time. The phase coexistence state is quantified by investigating the global and local properties of both the phases of the system during its transition. This experiment is the first of its kind where phase coexistence has been studied by changing the local sheath structure rather than changing the discharge properties.

\par The paper is organised as follows: in the next section, Sec.~\ref{expt}, we describe the experimental arrangement in detail. In Sec. \ref{result}, we discuss the experimental results on the formation of the dust crystal and phase coexistance state. In Sec. \ref{discussion} we discuss our experimental results, as well as the results of others, and how they should be interpreted given the available theoretical mechanisms. A brief concluding remark is made in Sec.~\ref{conclusion}.
%%%%%%%%%%%%%%%%%%%%%%%%%%%%%%%%%%%%%%%%%%%%%
\section{Experimental arrangement\label{expt}}
%%%%%%%%%%%%%%%%%%%%%%%%%%%%%%
\begin{figure*}
  \includegraphics[scale=0.5]{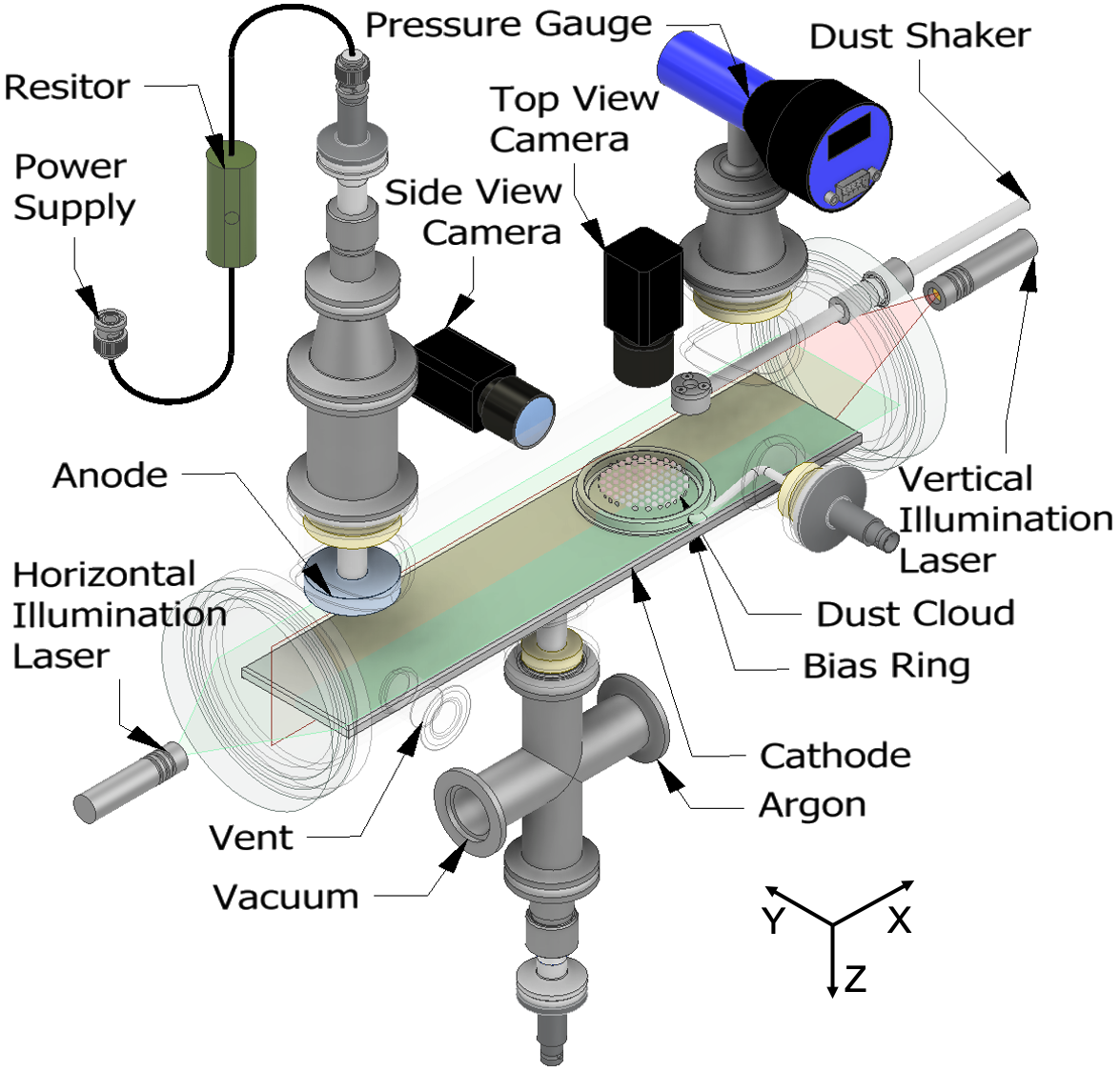}
  \caption{Schematic diagram of the experimental arrangement.}
  \label{fig:fig1}
\end{figure*}
%%%%%%%%%%%%%%%%%%%%%%%%%%%%%%
%%%%%%%%%%%%%%%%%%%%%%%%%%
\begin{figure*}
  \includegraphics[scale=0.6]{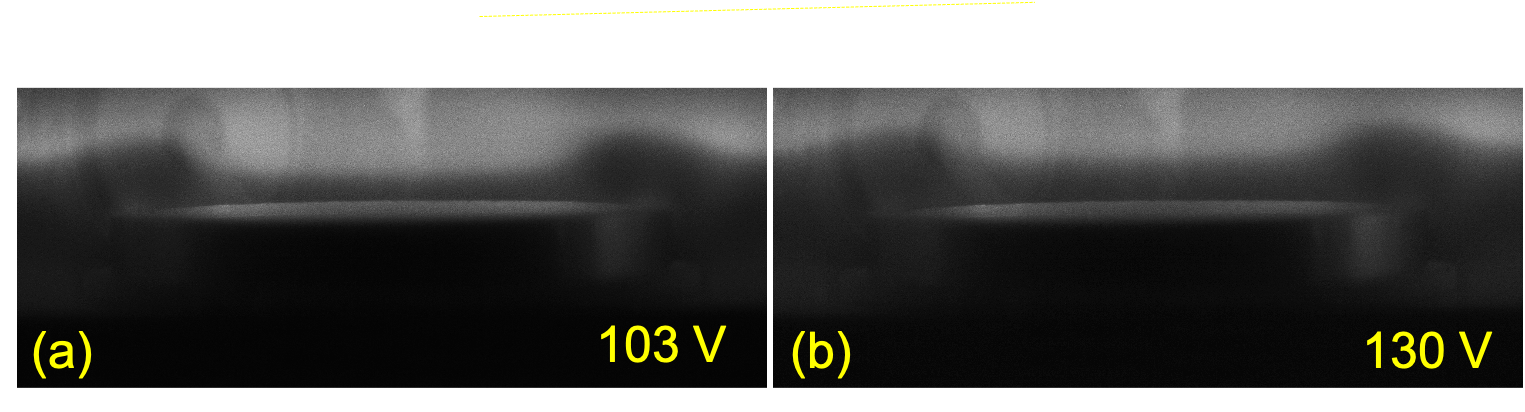}
  \caption{Photograph of the sheath structure inside the biased ring for different bias voltage (a) 103 V and (b) 130 V.}
  \label{fig:fig2}
\end{figure*}
%%%%%%%%%%%%%%%%%%%%%%%%%%
The experiment has been performed in a newly built linear dusty plasma experimental device (LDPEx). The schematic of the experimental setup is shown in Fig.~\ref{fig:fig1}. This is a very versatile and handy device which has the capability to create a stable, spatially extended large sized DC complex plasma crystal over a range of discharge conditions. This table top device is made of a borosilicate glass tube 33 cm long and 3 inch (7.6 cm) in inner diameter. The device has a number of standard axial and radial ports for ease of accessibility. Generally, for a discharge configuration in a DC system with symmetric electrode and equal electrode areas, the flux of electrons at the anode becomes equal to the flux of ions at the cathode. This creates significant ion heating on the dust particle and decreases the possibility of forming the dust crystal. Therefore, in the present experimental setup we have utilized the concept of using an asymmetric electrode configuration with a disc shaped stainless steel (SS) anode of 1 in. (2.5 cm) diameter and a long SS plate of 30 cm$\times$6 cm$\times$0.6 cm which serves as the cathode. The anode was inserted through a KF-40 vertical port and the cathode was placed horizontally at a distance of approximately 6 cm from the anode. At this distance, we have observed the most stable plasma discharge for the range of discharge parameters. The asymmetric electrode configuration for DC discharge dusty plasma experiments was first utilized by Jaiswal \textit{et. al} \cite{jaiswal2015} in the DPEx device and adapted in recent publications \cite{jaiswal_2016_PSST, swarnima_pop, saravananPSST_2021}. The asymmetry in the electrode configuration with a longer cathode area lowers the ion flux at the cathode as compared to the electron flux at the anode resulting in a reduced ion heating effect on the dust particles. This reduction in ion heating is a function of the ratio of anode to cathode areas \cite{saravananPSST_2021}. Therefore, the benefit of using a DC discharge to study dusty plasma crystals is that it provides better control of ion drift effects on the dust cloud. In contrast to previously reported experiments where a tray was used instead of a plate as a medium to implement the reduced ion heating, we have not observed burnt areas on the cathode surface even after running the experiments for a prolonged time. Moreover, this arrangement provides extra visibility from the sides and the small diameter of the chamber helps confine the particles better when making a long dusty plasma cloud. In the previous generation of asymmetric DC systems, the pumping and gas arrangement were causing a directed motion of neutrals along the axis of the cathode where the dust cloud forms. As a result, maintaining a flow free stable cloud was a challenge and was only possible by maintaining the pumping speed and gas flow rate. Additional confinement (side bending, multiple confining strips) was also required to hold the dust cloud in place. In this new system, we have resolved that issue by connecting the gas and pumping port adjacent to each other underneath the cathode. The result is a diffusive plasma and a large stable dust crystal without manipulating the gas flow or pumping speed. We use PEEK insulation at every port to remove the hollow cathode discharge effect. 

\par A rotary pump provides a stable base pressure of $10^{-3~}$ mbar in the vacuum chamber. To remove any kind of impurity, Argon gas was flushed several times in the chamber and pumped down to base pressure. Finally, the working pressure was set to 112 mTorr (15 Pa). A DC glow discharge plasma was formed in between the anode and cathode by applying a discharge voltage in the range of 350-400 V. The discharge current of 2-5 mA was estimated by measuring the voltage drop across a 2 k$\Omega$ resistor, which was connected in series with the power supply. Mono-disperse melamine-formaldehyde spheres of diameter $7.14\pm 0.06~\mu$m were introduced into the plasma by shaking a dust dispenser which was inside the vacuum chamber. These particles acquire negative charge by collecting more electrons than ions and get trapped in the plasma sheath boundary above the grounded cathode inside the biasing ring. In this levitating condition, the vertical component of the sheath electric field provides the necessary electrostatic force to the particles to balance the gravitational force. The particle-cloud is illuminated by a horizontally expanded thin laser sheet (532 nm, 100 mW) which was sufficiently constricted vertically to study an individual layer of the dust-cloud. A second laser (650 nm, 100 mW) was used to illuminate the particle in the vertical direction. A top view camera  with a green filter ($532\pm 4~$nm) attached was used to record the images in the horizontal (XY) plane, whereas a side view camera with a red filter ($650\pm 4~$nm) attached captured images in the vertical (XZ) plane. The top and side view cameras captured the images at 99 fps and 160 fps, respectively, with a resolution of 25 $\mu$m/pixel. The images were stored in a high speed computer. The sequence of images were analyzed using a Python-based tracking software (Trackpy) \cite{trackpy} and ImageJ software and the results were plotted using Python and MATLAB software.

\par The main difference in our present study as compared to previous experiments on phase coexistence is that we do not vary the discharge parameters to alter the dust cloud. Instead, the potential of a biased ring made of aluminum with an inner diameter of 5 cm and a height of 8 mm is used as a control parameter. %we have altered the dynamics of plasma crystal by changing the local sheath structure around a confining ring. 
The radial sheath electric field due to the circular confining ring is responsible for the radial confinement of the dust particles against their mutual Coulomb repulsive forces and allows the formation of a strongly coupled 3D dust crystal. The sheath around the ring also aids in the vertical confinement of the dust layer along with the cathode sheath. %The strength of this trap depends on the plasma parameters defining the sheath thickness and the electric field strength.
The confinement ring in this experiment was kept isolated from the cathode by a ceramic base as shown in Fig.~\ref{fig:fig1}. The potential of the ring is varied by applying an external DC bias which manipulates the sheath around the confinement ring in both the vertical and horizontal directions (Fig.~\ref{fig:fig2}). This affects the levitation height and the overall behaviour of the dust cloud. As can be seen in Fig.~\ref{fig:fig2}, the confinement strength becomes stronger by increasing the bias voltage from 103 V to 130 V with the sheath structure extending in the vertical direction. As we have measured, and also reported by Swarnima \textit{et al.} \cite{swarnima_pop}, the bulk plasma properties are not affected in this configuration. Hence, by using this technique, one is able to tune the vertical and radial confinement without altering the discharge (or plasma) conditions.  

\section{Experimental Results \label{result}}
%%%%%%%%%%%%%%%%%%%%%%%%%%%%%%%%%%%%%%%%%
\begin{figure*}
  \includegraphics[scale=0.47]{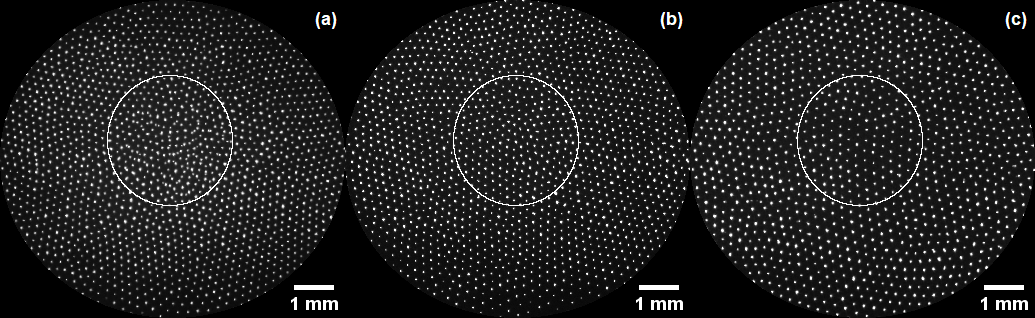}
  \caption{Representative images for the top view of the dust cloud for bias voltage (a) 103 V, (b) 110 V, and (c) 140 V. The white circular section represents the region of structural transition.}
  \label{fig:fig3}
\end{figure*}
%%%%%%%%%%%%%%%%%%%%%%%%%%%%%%%%%%%%%%%%%
In rf plasma the behaviour of the dust cloud depends mainly on two parameters, rf power and background pressure, but also depends on the number of dust layers formed. A larger number of layers corresponds to a more disordered crystalline structure. The layer which is closer to the electrode melts first and shows more randomized particle motion whereas the uppermost layer remains in a specific state for a longer period of time. On the other hand, in DC systems it was observed that the transition depends on discharge voltage and background pressure. However, a small variation in discharge voltage changes the plasma condition abruptly, therefore a clear variation in system dynamics is difficult to realize. Pressure is another important parameter in observing the melting and phase coexistence. Both the discharge voltage and pressure affect the plasma parameters globally. Therefore, we fixed the discharge voltage at 380 V and pressure at 15 Pa to keep the plasma condition as stable as possible and performed the experiment by varying the ring bias voltage. In this configuration the sheath structure inside the bias ring changes (see Fig.~\ref{fig:fig2}) and affects the particle cloud. At this voltage and pressure, a stable bilayer crystalline structure was formed when we set the ring bias voltage at 140 V. A snapshot of the Coulomb crystal is shown in Fig.~\ref{fig:fig3}(c). It is clearly visible that the dust particles arrange into a regular hexagonal lattice with nearly uniform particle spacing and good translational periodicity which represents that the system has long range ordering. The top layer was unaffected by the layer beneath it %and most of the crystal points do not show any deviation except a random motion around their mean positions with a average kinetic energy of around 0.04 eV. 
and the structure is almost stationary with a small thermal fluctuation around the equilibrium position that we have checked from the overlapped image of 100 consecutive frames. The average kinetic temperature in this state was very low, around 0.04 eV. This indicates that the particles are almost aligned in the vertical direction and the lower layer does not make any significant effect on the upper layer when the ring bias voltage is higher (140 V). In this configuration the sheath structure is extended to 4 mm above the confining ring.

\par By reducing the bias voltage the sheath structure compresses vertically and is much closer to the confinement ring as shown in Fig.~\ref{fig:fig2}(a). We observed that the core of the dust crystal began oscillating at 110 V (Fig.~\ref{fig:fig3}(b)) and transformed into a fluid phase when the voltage was further reduced to 103 V as shown in Fig.~\ref{fig:fig3}(a). The region inside the circle represents the location where changes in the particle dynamics were taking place as a function of bias voltage. The particles outside the circle maintain their crystalline behaviour irrespective of the change in bias voltage.

\par In order to better understand the phase coexistence of the dust cloud with changing ring bias voltage, we performed a variety of diagnostic tests on the experimental data by calculating the pair correlation function, the local bond order ($\Psi_6$), and estimating the dust temperature. These parameters give information on global as well as local structural properties of the cloud. Below we provide a detailed description of the results of our analysis. We kept the discharge voltage and pressure at a constant 380 V and 15 Pa throughout the experiments while changing the ring bias voltage from 100 to 140 V.

%%%%%%%%%%%%%%%%%%%
\begin{figure}
  \includegraphics[scale=0.45]{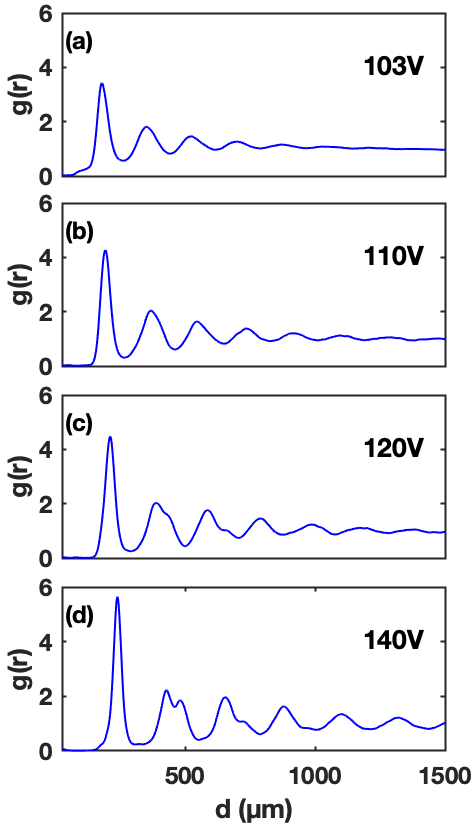}
  \caption{Pair correlation function of the selected region of the cloud as a function of ring bias voltage.}
  \label{fig:fig4}
\end{figure}
%%%%%%%%%%%%%%%%%%%

\par
\begin{figure}
  \includegraphics[scale=0.4]{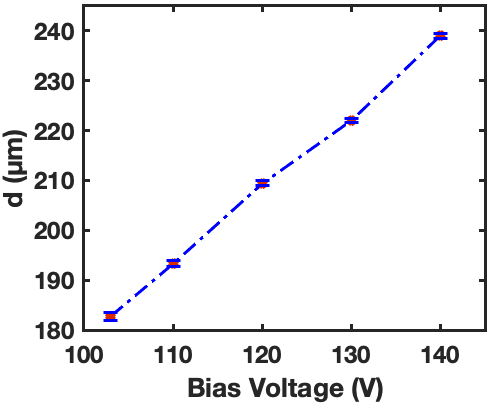}
  \caption{Inter-particle distance variation with respect to ring bias voltage.}
  \label{fig:fig5}
\end{figure}

 The radial pair correlation function, $g(r)$, is defined as a measure of the probability of finding a particle at a distance $d$ from the reference particle and is calculated by directly measuring the average distance between particles. Fig.~\ref{fig:fig4} presents the radial pair correlation function which delineates the global system properties in terms of the degree of long range order of the distribution of particles. In this calculation, 350 frames were used for averaging. As we focus on understanding how the positional ordering of the structure is changing as a function of ring bias voltage, $g(r)$ vs. $d$ of the structure was calculated for the region inside the circle (shown in Fig.~\ref{fig:fig3}). As shown in Fig.~\ref{fig:fig4}(d), for bias of 140 V, the positional order of the structure is preserved up to the sixth nearest neighbour, revealing the presence of a long range ordering and a nearly perfect crystal. This is also confirmed from the clear splitting of the second peak which is a signature of an ideal hexagonal crystal \cite{jaiswal_melting}. With the decrease in bias voltage, the peaks become shorter and flattened, showing the increasing disorder of the cloud. Additionally, the number of peaks and the length to which ordering is observable becomes decreased. The splitting of the second peak and the signature of long range ordering is visible up to 120 V. At 110 V, (see Fig.~\ref{fig:fig4}(b)) the splitting in the second peak disappeared and the peaks become more flattened. This indicated that the phase state of the particle changes and it tends towards the liquid state with a primary peak followed by quickly descending peaks. At 110 V, the bottom particles started to oscillate as manifested by the disorder in the particle's position in the central region in Fig.~\ref{fig:fig3}(b) which illustrates that this is the threshold value below which the particle cloud melts in the core region. At 103 V, visualized in Fig.~\ref{fig:fig3}(a), the pair correlation function shows a signature of a nearly fluid state. The primary peak is very small with very fast descending second or third peaks. These pair correlation function measurements indicate a transition from an ordered solid structure to a liquid phase in the middle of the cloud at reduced ring bias voltage. It is worth mentioning that the region outside of the circle does not show the same signature in the pair correlation function which shows that the periphery remains in crystalline phase irrespective of the changing ring bias voltage (changing sheath structure). The pair correlation function also provides a good estimate for the mean inter-particle spacing from the position of the first peak. The mean inter-particle distance vs. ring bias voltage, which varies from 182 - 240 $\mu$m for 103 - 140 V, is plotted in Fig.~\ref{fig:fig5}. With a large ring bias voltage, the inter-particle gap between the particles increases which may be due to the vertically expanding sheath structure (see Fig.~\ref{fig:fig2}). Therefore, the particles formed a distinct two layers at a high ring bias voltage.
%%%%%%%%%%%%%%%%%%%%%%%%%%%%%%%%%%
%%%%%%%%%%%%%%%%%%%%%%%%%%%%%%%%%%
%%%%%%%%%%%%%%%%%%%%%%%%%%%%%%%%%%
\begin{figure}
  \includegraphics[scale=0.35]{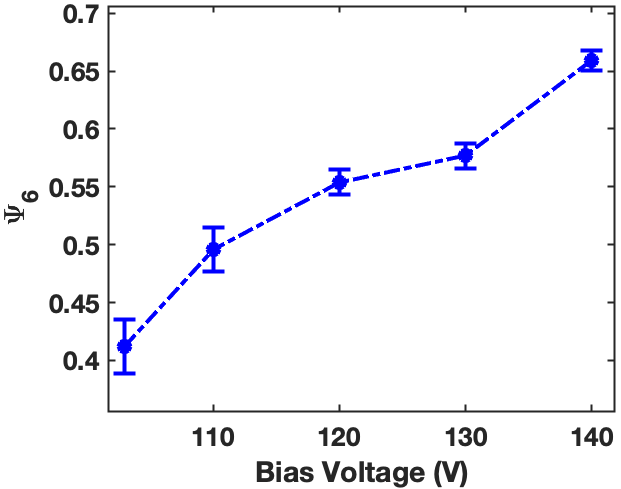}
  \caption{Variation of the average bond order parameter $\Psi_6$ with ring bias voltage.}
  \label{fig:fig6}
\end{figure}
%%%%%%%%%%%%%%%%%%%%%%%%%%%%%%%%%%
\par We further verify our understanding by measuring the variation in structural dynamics of the particle cloud locally. We calculated the local bond order parameter as a function of bias voltage. This quantity is defined locally and measure the structural properties of the lattice directly at the respective particle positions \cite{christina_2007, christina_thesis}. A bond order parameter, $\Psi_6$, examines the lattice in terms of the local orientational order of particles \cite{christina_thesis, coudel_2018}. For a given particle $k$, $\Psi_6$ is defined as:

\begin{equation}%\label{eq:eq1}
\Psi_6(k) = \frac{1}{N}\left|\sum_{n=1}^{N}e^{6i\theta_{kn}}\right|,
\end{equation}

where $N$ is the number of nearest neighbors and $\theta_{kn}$ is the angle of the bond between the $k_{th}$ and $n_{th}$ particles with respect to the x-axis. For an ideal hexagonal structure, the bond order parameter is a maximum ($\Psi_6 = 1$) hence it is a good measure of the deformation of a cell from the perfect hexagon. Lattice sites with nearest neighbor bond angles deviating from 60$^\circ$ will decrease $\Psi_6$. The same occurs for lattice sites with a number of nearest neighbors other than six. For calculating the bond order parameter, we first identify the particle locations and find the nearest neighbors of all the particles using Delaunay triangulation. Next, for every particle $k$, we calculate a vector going to its nearest neighbours $n$ and then calculated the angle of the vector $kn$ with respect to the x-axis. The variation in average bond order parameter, $\Psi_6$, with ring bias voltage is plotted in Fig.~\ref{fig:fig6}. The mean $\Psi_6$, which is obtained by calculating $\Psi_6(k)$ for each unit cell and then taking the average over all the cells, yields information of the overall orientational ordering of the system. We have chosen the same circular region for this analysis which we have chosen for the calculation of pair correlation function. The local order of the particle cloud is $\sim$ 0.68 at 140 V. $\Psi_6(k)$ shows a decreasing trend with reducing ring bias voltage. The minimum value of $\Psi_6$ is $\sim$ 0.4 for 103 V, which is close to the values reported for the melting dynamics in rf plasmas \cite{christina_thesis}. The $\Psi_6(k)$ measurements corroborate that greater structural order is achieved with an increase in ring bias voltage. It also signifies that particles tend to the fluid state below 110 V which confirms that below a threshold voltage, the system manifests phase coexistence with a fluid state in the core and a crystalline state at the periphery.
%%%%%%%%%%%%%%%%%%%%%%%%%%%%%%%
\begin{figure}
  \includegraphics[scale=0.7]{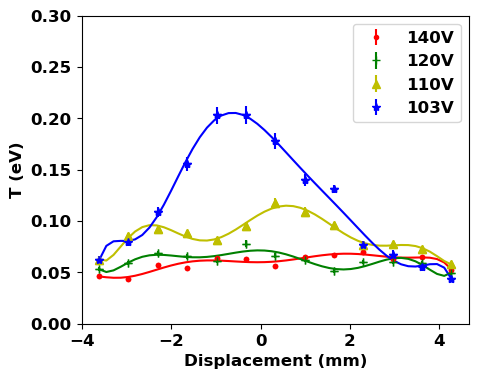}
  \caption{Local temperature measurement with the dust cloud as a function of ring bias voltage. }
  \label{fig:fig7}
\end{figure}
%%%%%%%%%%%%%%%%%%%%%%%%%%%%%%%%%%

\par The thermodynamic properties of the system are further determined by measuring the kinetic temperature of the dust particles. For the experimental measurement of the kinetic temperature a velocity distribution function was calculated. For better statistics, a video sequence of 300 frames was chosen where a Python based particle tracking analysis library (Trackpy) was used to track the coordinates of the individual particle which was used to construct the displacement and then velocity vector fields. In our experiment, the velocity distribution function was found to be Maxwellian. By taking local temperature measurements from the periphery through the core of the cloud (see images of Fig.~\ref{fig:fig3}), changes in kinetic temperature are visualized as the system begins to melt at the core. Therefore, the entire cloud is divided into multiple segments along the length of the crystal and the velocity distribution was calculated for each segment. The kinetic dust temperature was then calculated from the width of the measured distribution for each segment of the cloud by using the formula $E =\frac{1}{2}m \langle v^2_{x,y} \rangle = \frac{1}{2}k_B T_{x,y}$.  

%%%%%%%%%%%%%%%%%%%%%%%%%%%%%%%%%%
\begin{figure*}
  \includegraphics[scale=0.475]{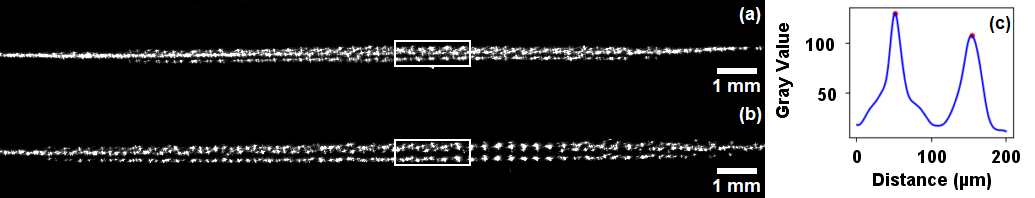}
  \caption{Representative image taken from the side camera for a ring bias voltage of (a) 103 V and (b) 140 V with (c) a corresponding average vertical intensity profile for 140 V.}
  \label{fig:fig8}
\end{figure*}
%%%%%%%%%%%%%%%%%%%%%%%%%%%%%%%%%%

\par Fig.~\ref{fig:fig7} shows the measured kinetic temperature of the dust particles in the axial direction as we change the ring bias voltage form 103 V to 140 V. At 140 V, where the system exhibits a crystalline state, the temperature remains uniform along the axial extent and it is close to 0.04 eV. At 120 V, the temperature variation is largely the same as at 140 V. A small increase appears at the center of the cloud while maintaining a uniform temperature at the periphery when the ring bias voltage reached the threshold value of 110 V. Below this threshold, the temperature increases significantly in the core region of the cloud whereas it is unchanged at the edges of the cloud compared to higher bias voltages. This confirms that the melting at the core begins around 110 V. Additionally, it further proves that the system maintains a state of phase coexistence, since a uniform temperature is maintained at all bias voltages around the periphery of the cloud.

%%%%%%%%%%%%%%%%%
\begin{figure}
  \includegraphics[scale=0.6]{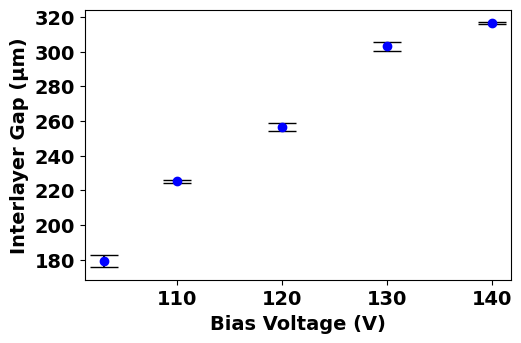}
  \caption{Interlayer gap vs. ring bias voltage. }
  \label{fig:fig9}
\end{figure}
%%%%%%%%%%%%%%%%%

\par To investigate mechanisms that could be responsible for the melting of the dust particles in the core, the side images of the dust cloud are analyzed. Representative images recorded from the side camera are shown in Fig.~\ref{fig:fig8}(a) and (b). {Fig.~\ref{fig:fig8}(c) shows the average intensity profile corresponding to the rectangular region marked in Fig.~\ref{fig:fig8}(b). The difference between the peaks is used to calculate the gap between two layers. When the ring bias voltage was 103 V, see Fig.~\ref{fig:fig8}(a), the bi-layer region compressed vertically and horizontally. This results in localization of the particles in the core and an increased monolayer periphery. This decreased bi-layer region also leads to a smaller interlayer gap. An increase in bias voltage to 140 V, as visualized in Fig.~\ref{fig:fig8}(b), shows a distinct bi-layer system that expands much closer to the peripheral region of the cloud. This expansion leads to a larger, more defined interlayer gap. The visual change in the interlayer gap with increasing bias voltage is confirmed by the results presented in Fig~\ref{fig:fig9}. With increasing bias voltage, the interlayer gap increases. 

\par The difference in the overall structure for varying bias voltage is further explained by referring to the sheath structure around the ring seen in Fig.~\ref{fig:fig2}. The sheath structure compresses in the vertical direction more than the horizontal direction with decreasing bias voltage and moves closer to the cathode at 103 V. The reduction of the sheath both horizontally and vertically with decreasing bias voltage causes the bi-layer region to decrease along with the interlayer gap. A reduction in sheath height brings the layers closer together and when the interlayer gap becomes comparable/smaller than the interparticle distance, the particles in the bottom layer repel the top layer which starts both in-plane and vertical fluctuations in the particles. The fluctuation grows with reducing bias voltage and accompanies melting of the core.

%%%%%%%%%%%%%%%%%%%%%%%%%%%%%%%%%%
\begin{figure}%[ht]
  \includegraphics[width=.9\columnwidth]{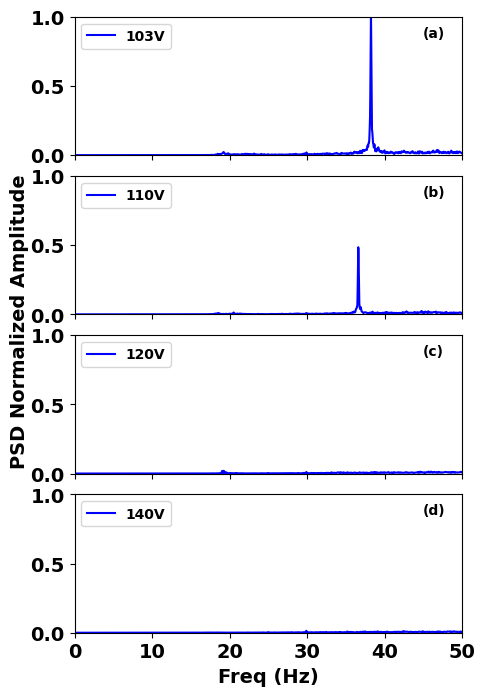}
  \caption {Power density spectra of horizontal particle oscillations as a function of frequency for 103 V, 110 V, 120 V, and 140 V ring bias voltage.}
  \label{fig:fig10}
\end{figure}
%%%%%%%%%%%%%%%%%%%%%%%%%%%%%%%%%%
%%%%%%%%%%%%%%%%%
\begin{figure}
  \includegraphics[width=.9\columnwidth]{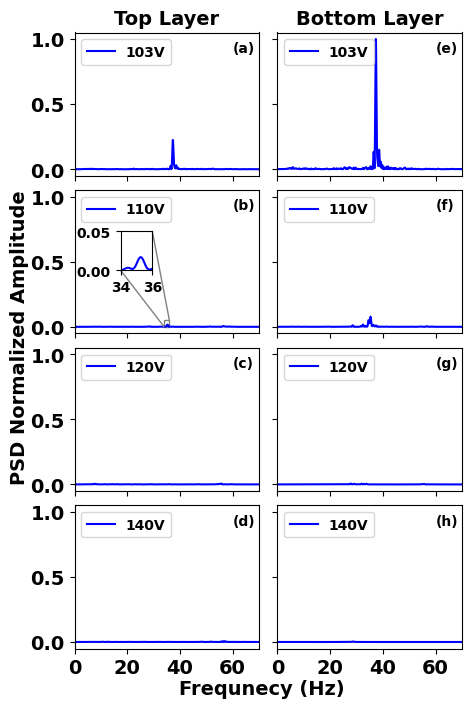}
  \caption{Normalized power spectral density for vertical oscillation as a function of frequency for various ring bias voltages for the (a-d) top and (e-h) bottom layer of the bi-layer cloud.}
  \label{fig:fig11}
\end{figure}
%%%%%%%%%%%%%%%%%
\par To quantify our observations, we looked at both the particle's in-plane and interlayer oscillations as the crystalline structure begins melting into a phase coexistence state. The frequency of oscillation for varying bias voltage is calculated using the Power Spectral Density (PSD). For in-plane particle oscillations, the PSD as a function of frequency plots are shown in Fig.~\ref{fig:fig10} for various voltages. The individual particle coordinates were obtained using ImageJ and the trajectories were found using Trackpy, which utilizes the Crocker-Grier algorithm \cite{Crocker}. The 25 particles with the longest trajectories confined within the defined fluid region were selected for the spectral analysis. The particles used for analysis were tracked across a minimum of 750 frames and the drift subtracted using an asymmetric least squares smoothing algorithm. Butterworth high-pass filter was used to suppress the DC bias. The PSD plots were then obtained by performing a Fast Fourier Transform FFT of the signals for all the selected particles. Finally, the 25 PSD spectra of individual particles were superimposed to get the final spectra. As seen from Fig.~\ref{fig:fig10} (c) and (d), the in-plane oscillation shows no defined oscillation frequency peak until we reduced the bias voltage to the threshold value. At 110 V, the spectra shows an oscillation frequency peak around 36 Hz, and at 103 V the peak was at a frequency of around 38 Hz. A significantly smaller peak at 19 Hz is observed across all bias voltages and is believed to be the frequency of natural oscillation of the particle cloud. The results confirm that once the threshold bias voltage is reached, the particles in the core begin to oscillate with growing amplitude with reducing bias voltage to 103 V and cause the melting within the core of the cloud. We tracked the fluctuations in position of the bi-layer structure in time. This fluctuation was calculated by analyzing the vertical average intensity profile for a 75-pixel horizontal region in the cloud and then fit with a smoothing spline. An example of this intensity profile is shown in Fig.~\ref{fig:fig8}(c) for 140 V bias. Since we are interested in the transition of the system from a crystalline to a phase coexistence state, the center of the melting region was chosen for analysis. Each profile contained two local maxima whose distance represented the interlayer gap for each corresponding frame. The positions of both peaks were tracked for 300 frames. To calculate the PSD for the top and bottom layer we followed the steps similar to those made to calculate the PSD of the in-plane osciallations and described above. The mean was subtracted and passed through a Butterworth high-pass filter to suppress the DC bias. Then, the FFT of the resultant signal was performed. All spectra for both layers were normalized using the amplitude of oscillation corresponding to the 103 V bias of the bottom layer. The resultant PSD that is shown in Fig.~\ref{fig:fig11} indicates that the interlayer frequency of oscillation when the structure is in a state of phase coexistence varies from 35-37 Hz for both layers. Once the bias voltage reaches 120 V, the interlayer oscillation becomes so sporadic that there is no longer a detectable frequency. This result confirms that at the threshold bias voltage where the layers are in close proximity to each other, both layers begin to show fluctuations. The oscillation of the bottom layer shows a significantly higher amplitude than that of the top layer for each corresponding bias voltage. The fluctuation in the bottom layer affects the particles in the primary top layer. The amplitude of each layer grows with reducing bias voltage. 
%which initiate the melting in the core of the cloud. 
Interestingly, the frequency of oscillation for in-plane and interlayer are comparable to each other indicating that they are in resonance once we approach the threshold value of ring bias voltage.  

%%%%%%%%%%%%%%%%%
\begin{figure}
  \includegraphics[width=1.0\columnwidth]{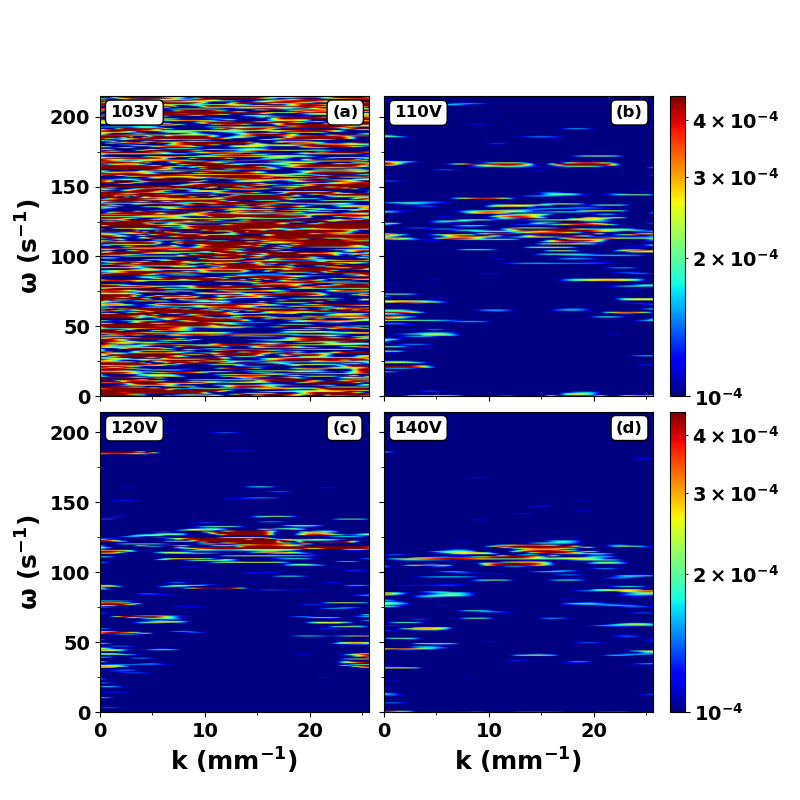}
  \caption{Evolution of the in-plane phonon spectra of the particle cloud in the region of structural transition. The logarithmic color is the intensity of the spectra in arbitrary units.}
  \label{fig:fig12}
\end{figure}
%%%%%%%%%%%%%%%%%
\par In Fig.~\ref{fig:fig12}, the phonon spectra of the particle cloud in-plane motion in the region of structural transition are depicted. The figure illustrates the melting transition of the crystal as the bias voltage decreases. At a bias voltage of 140 V, seen in panel (d), the phonon spectrum has features typical of a solid. As the voltage is reduced to 110 V, these features start to broaden. By 103 V, which corresponds to the liquid state, these features are largely absent, resulting in an almost uniform field. Importantly, the spectra do not display the features characteristic to the MCI. Although we don't possess the out-of-plane branch of the phonon spectra to directly observe the in-plane and out-of-plane branch intersection - a hallmark of the MCI - our in-plane spectrum distinctly lacks the horizontal line of “hot spots” that arise from such intersection \cite{lenaic}.
\section{Discussion\label{discussion}}
To draw a complete understanding of the cause of phase coexistance, it is important to investigate the mechanisms for melting of dust crystals, some of which were mentioned in the introduction. There are two main mechanisms, the MCI \cite{coudel_review} and the Schweigert instability \cite{schweigert3} shown to be responsible for melting of the plasma crystal. Each has particular relevance given the experimental conditions and a unique mechanism for explaining this behaviour. The Schweigert instability \cite{schweigert3} was shown to be applicable for multilayer crystals while the MCI \cite{coudel_review} is relevant in the case of monolayer crystals. Importantly, both instabilities explain melting as the neutral pressure is varied; below a critical value the gas friction cannot damp out the instability which grows and melts the crystal. 

\par Experimental evidence for the Schweigert instability was presented by several authors \cite{schweigert3,ichiki}. For example, in the paper by Ichiki \textit{et al.} \cite{ichiki} a single particle was placed beneath a monolayer crystal. This particle is vertically aligned to a top particle due to plasma wake effects. At a critical pressure, the horizontally oscillating vertical pair forces all particles in the upper layer to also oscillate at a particular frequency. The melting proceeds in two stages, depending on neutral pressure; first the crystal transitions to a hot crystalline state and further reduction of the neutral pressure completely melts the entire crystal. Vertical coupling at a higher frequency was not observed. They also found that in the absence of a particle underneath the primary top layer the crystal did not melt because the Schweigert instability is relevant only for two or more layer systems \cite{ichiki,schweigert1}. On the other hand as explained in detail by Cou{\"e}del \textit{et al.} \cite{coudel_review}, the MCI requires the existence of a hybrid mode, which arises when the in-plane (horizontal) dust lattice mode and the out-of-plane (vertical) mode intersect. In addition to existence of the hybrid mode, the MCI requires two conditions: the first is based on the amount of gas damping in the system and the second determines the vertical confinement. The MCI also requires a certain value of particle density. When the hybrid mode is present and the above conditions are met, the entire monolayer crystal melts. A further reduction in the vertical confinement leads to multilayer systems.

\par This understanding implies that both the MCI and Schweigert mechanisms are not applicable to our system because we have a multilayer system and absence of horizontal line of hot spots (this excludes MCI). Since we observe clear vertical oscillations in both layers this excludes the Schweigert instability. Moreover, both mechanisms are not applicable to our system because they have a threshold in damping strength beyond which the crystal melts. At a constant neutral pressure (i.e. damping strength), changes in the crystal structure have been shown in our work to result from the strength of the confining potential which indicate that the threshold on the damping strength is irrelevant in our case. However, these mechanisms may be contributing some physical signatures in our system in addition to some other effects which would be responsible for phase coexistance. In order to discriminate between the various mechanisms for melting and phase coexistence, both the in-plane oscillations as well as the vertical oscillations must be investigated. This rigorous analysis was lacking in some of the recent experiments done by Swarnima \textit{et al.} \cite{swarnima_pop, hariprasad_phase} where they studied the phase coexistence in a monolayer system by lowering the gas pressure, as in the work by Cou{\"e}del \textit{et al.}\cite{coudel_review} 

\par In this work we further explored other possible mechanisms that could be responsible for the phase coexistance in our system due to changing confining potential strength (changing sheath structure). This was done by investigating the role of key parameters of the dusty plasma system. We have found that the interlayer gap and interparticle distance both decrease as shown in Fig.~\ref{fig:fig9} and Fig.~\ref{fig:fig5}, respectively for decreasing ring bias voltage. When these two lengths become comparable, we observe phase coexistence. The interlayer gap becomes small enough that the bottom layer particles exert an influence on the upper layer particles. The screening length, which establishes the influence of a dust particle to its surroundings, is generally several times that of the ion Debye length. Nikolaev and Timofeev \cite{screening} found that the average screening length is roughly 8 times that of the ion Debye length, ($\lambda_{Scr} \approx 8\lambda_{Ion} $, where $\lambda_{Ion} = \sqrt{\frac{\epsilon_{0} k_{B} T_{i}}{n_{e} e^{2}}}$) when inter-particle potentials were considered as Yukawa type. Here, $\epsilon_{0}$ is the permittivity of free space, $k_{B}$ is the Boltzmann constant, $T_{i}$ is the ion temperature, $n_{e}$ is the electron density, and $e$ is the elemental charge. For  our experimental values of $n_{e}=2 \times 10^{15}$ m$^{-3}$ and $T_{i} = 300$ K (taken to be the room temperature), the screening length is 214 $\mu$m. For a ring bias of 110 V and 103 V, the interlayer gap and interparticle distance are comparable to or below the screening length which establishes that the particle's behavior becomes quite complicated. In this case, we observe the phase coexistence because the two layers are so close to each other that they affect the other layer. When this occurs, the particles in each layer become close enough to repel each other due to their negative charge which starts vertical and eventually in-plane oscillations. The amplitude of these oscillations grows with reducing bias voltage which causes melting of the core. When the bias voltage was reduced the layers are compressed in both the vertical and horizontal directions due to the combined confining influence of the ring (radial confinement) and the cathode (vertical confinement). Therefore, the movement of the layer underneath the main layer is restricted to a limited spatial region in the core which undergoes horizontal and vertical oscillations. This indicates that the melting and phase coexistence is partly due to effects from sheath confinement.

\par The PSD measurements in Fig.~\ref{fig:fig10} and Fig.~\ref{fig:fig11} clearly showed that the bottom dust layer affects the top layer and transfers energy. Both layers achieve the same oscillation frequency and the result is a crystal melted in the central region with a solid periphery. The dust particles in the lower layer are affected by the ion wake effect as well. Recent work on simulations by Matthews \textit{et al.}\cite{ion_wake_sim} of the ion wake effect show that the potential sphere, relative to the average ion potential, of the particle pair changes to be nearly equal for two dust particles under some circumstances. In this case, the two particles would have a stronger repulsive force which could force the two particles away. This attractive and repulsive oscillation due to charge fluctuations can melt the upper layer as we explained above. Also, when the bottom particles vertically oscillate within the wake, they can become decharged significantly \cite{ion_wake_sim}. Our observations of the vertical oscillatory behavior suggest that these effects must also be considered when explaining phase transitions and phase coexistence. A self consistent simulation exploring these effects will be conducted in future work to form a more complete understanding of phase coexistence.

\section{Conclusion \label{conclusion}}
\par In conclusion, we have presented an experimental investigation on the structural transition of a dust crystal in a DC glow discharge plasma that results in solid-liquid phase coexistence while only changing the confinement potential. We have characterized the system by measuring the pair correlation function, local bond order parameter, and dust temperature. These measurements clearly show that the central core of the crystal is in a liquid state while the periphery remains crystalline when the confinement potential is reduced below a threshold value of 110 V. The phase coexistence results from the vertical oscillation of the bottom layer particles which affect the top layer particles at the confinement threshold. The confinement potential changes the sheath structure and the interlayer gap of the particles such that the interlayer gap and interparticle distance are comparable. When this occurs, the dust particles are close to (or within) the influence of the screening length which affects their trajectory. In this scenario, the particles experience increased energy transfer, charge fluctuations, and more chaotic behavior. Our work here has showcased that melting and phase coexistence in dusty plasmas can be realized in a different way than past work (where the pressure was the defining parameter) and that instabilities such as the Schweigert instability and mode coupling instability do not adequately explain the observations. More focused investigation of the plasma behavior and sheath characteristics are required to understand phenomena in dusty plasmas like we have observed here. Our findings may also prove useful for exploring other innovative modifications in various DC glow discharge devices to make them suitable for the study of dusty plasma crystals and related phenomena.  

\section{Acknowledgments}
S. J. acknowledges undergraduate students Parth Mehrotra and Siddhartha Mangamuri for contributions in the development of the analysis techniques. S. J. also acknowledges Eastern Michigan University for infrastructure support and Dr. Jeremiah Williams for infrastructure support and fruitful discussions. 

\bibliography{bibbly}

\end{document}